\documentclass[reprint,superscriptaddress,showpacs,amsmath,amssymb,aps,prl]{revtex4-1}
\usepackage{graphicx}
\usepackage{dcolumn}
\usepackage{bm}
\usepackage{hyperref}
\usepackage{subfigure}
\hypersetup{colorlinks=true, citecolor=blue, urlcolor=blue, linkcolor=blue}
\usepackage{amssymb}
\usepackage{color}
\usepackage{graphicx}
\usepackage{amsmath}
\usepackage{mathrsfs}
\usepackage{times}
\usepackage{subeqnarray}
\usepackage{cases}
\usepackage{bm}
\usepackage{diagbox}
\usepackage{booktabs}
\usepackage[table,xcdraw]{xcolor}
\usepackage{colortbl}
\usepackage{epstopdf}
\definecolor{tabcolor}{rgb}{.105,.410,.113}
\begin{document}

%\title{Trust and trustworthiness evolution based on Q-learning algorithm}
%\title{The evolution of trust and trustworthiness with Q-learning: the introspective paradigm}
%\title{The evolution of trust within the introspective paradigm: a Q-learning implementation}
%\title{The evolution of trust within the introspective paradigm: a reinforcement learning implementation}
%\title{Trust and trustworthiness evolution based on Q-learning algorithm}
%\title{The evolution of trust and trustworthiness: from the reinforcement learning perspective}
%\title{Explaining trust and trustworthiness: from the reinforcement learning perspective}
%\title{Decoding trust and trustworthiness: A reinforcement learning perspective}
\title{Decoding trust: A reinforcement learning perspective}

\author{Guozhong Zheng}
\affiliation{School of Physics and Information Technology, Shaanxi Normal University, Xi'an 710061, P. R. China}
\author{Jiqiang Zhang}
\affiliation{School of Physics and Electronic-Electrical Engineering, Ningxia University, Yinchuan 750021, P. R. China}
\author{Jing Zhang}
\affiliation{School of Physics and Information Technology, Shaanxi Normal University, Xi'an 710061, P. R. China}
\affiliation{College of Information Science and Technology, Donghua University, Shanghai 201620, P. R. China}
\author{Weiran Cai}
\email[Email address: ]{wrcai@suda.edu.cn}
\affiliation{School of Computer Science, Soochow University, Suzhou 215006, P. R. China}
\author{Li Chen}
\email[Email address: ]{chenl@snnu.edu.cn}
\affiliation{School of Physics and Information Technology, Shaanxi Normal University, Xi'an 710061, P. R. China}

\begin{abstract}
Behavioral experiments on the trust game have shown that trust and trustworthiness are universal among human beings, contradicting the prediction by assuming \emph{Homo economicus} in orthodox Economics. This means some mechanism must be at work that favors their emergence. Most previous explanations however need to resort to some exogenous factors based upon imitative learning, a simple version of social learning. Here, we turn to the paradigm of reinforcement learning, where individuals revise their strategies by evaluating the long-term return through accumulated experience. Specifically, we investigate the trust game with the Q-learning algorithm, where each participant is associated with two evolving Q-tables that guide one's decision making as trustor and trustee respectively. In the pairwise scenario, we reveal that high levels of trust and trustworthiness emerge when individuals appreciate both their historical experience and returns in the future. Mechanistically, the evolution of the Q-tables shows a crossover that resembles human psychological changes. We also provide the phase diagram for the game parameters, where the boundary analysis is conducted. These findings are robust when the scenario is extended to a latticed population. Our results thus provide a natural explanation for the emergence of trust and trustworthiness, and indicate that the long ignored endogenous factors alone are sufficient to drive. More importantly, the proposed paradigm  shows the potential in deciphering many puzzles in human behaviors. 
\end{abstract}

\date{\today }
\maketitle
\section{1. Introduction}\label{sec:introduction}
Trust and trustworthiness are central components of our human civilization~\cite{Hardin2002trust}, especially required in dealing with many pressing threats such as climate change, pandemic threats, international conflicts, energy crises etc. As ``a lubricant for social system"~\cite{Arrow1974limits}, trust can facilitate cooperation and contribute to the economic growth~\cite{Zak2001trust, Algan2013Trust}. A large-scale survey shows many of our human beings give a positive answer when facing the statement ``most people can be trusted", though a recent survey (see ``Using Data to Understand Our World")~\cite{Esteban2016Trust} shows that over the past four decades fewer people say they ``trust each other", which is an alarming signal. As fundamental questions, what is the mechanism for the emergence of trust and the associated trustworthiness, and under what conditions they are likely to sustain, have attracted considerable attention from different areas in the past decades.

Trust is the willingness of an agent (the trustor) to act in a way that benefits the other (the trustee) with the expectation that the trustee will return part of the profit afterwards. But, the trustor has no control over the trustee's action, the action of trust thus puts oneself in a vulnerable position. Without any mechanism to enforce the returns, the trustee tends to maximize her/his interest by simply walking away without reciprocation; with the backward induction, the trustor should not trust the trustee due to the expected loss of investment. That's the prediction from the assumption of~\emph{Homo economicus} in orthodox Economics~\cite{Simon1957Models, Francis1995Trust, Samuelson2005Economics}, where individuals are supposed to be self-interested, their behaviors are guided by the maximization of payoff.

A large number of laboratory experiments on the canonical model --- the trust game~\cite{Berg1995Trust}, however, reveal strikingly different observations~\cite{Berg1995Trust, Johnson2011Trust}. Across different countries with varied experimental protocols, the trustors are willing to send an average of $50\%$ of their initial endowments to trustees, and the trustees return an average of $37\%$ of the earnings back to trustors. Despite the variations in details, it turns out that we human beings remarkably prefer trusting others and are trustworthy to a large extent.

\begin{figure*}[ht]
\centering
\includegraphics[width=0.8\linewidth]{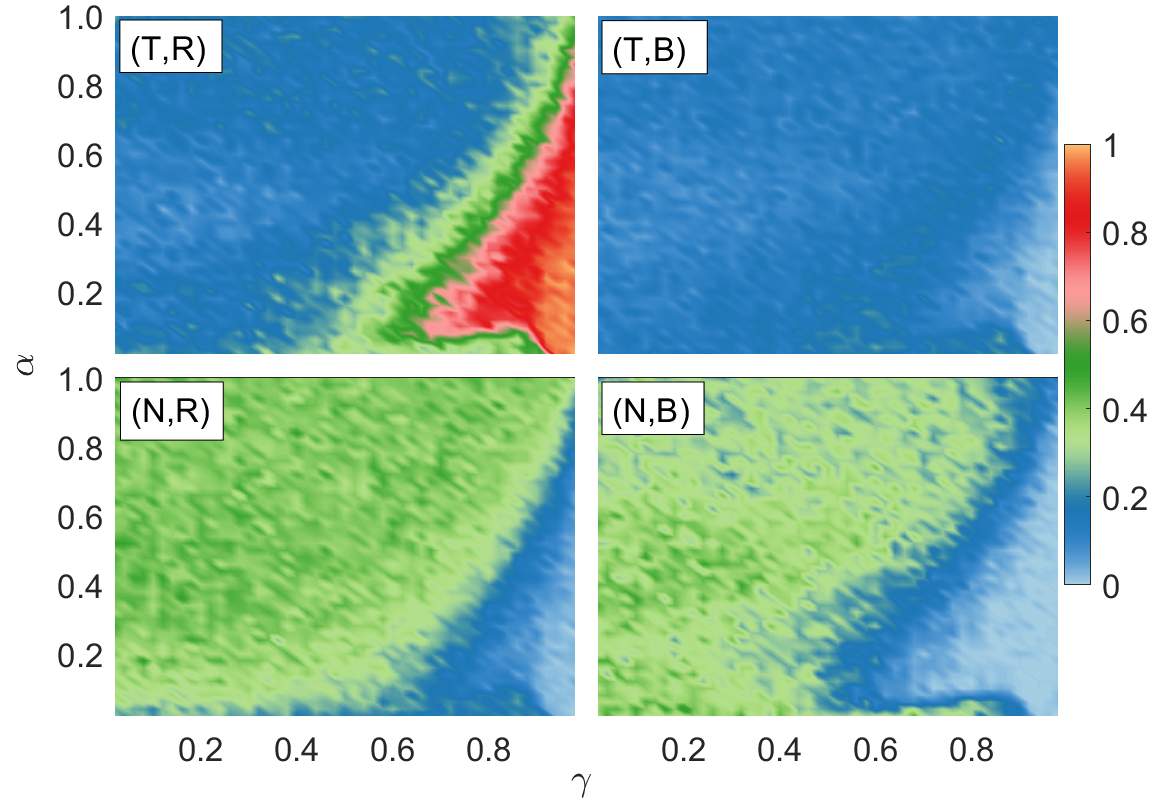}
\caption{%(Color online) 
\textbf{Emergence of trust and trustworthiness.}
The color-coded stationary fractions of the four strategies in the domain $(\gamma , \alpha)$ values with both ranged from $0$ to $1$. 
Here, e.g. the strategy combination TB means the player trusts the trustee when acting as a trustor, but chooses betrayal when acting as a trustee.
A large fraction of TR combination is seen in the corner of large $\gamma$ and small $\alpha$.
Each data is averaged over 100 realizations.
Other parameters: $\epsilon=0.01$, $g=3$, $x=1.0$, and $\omega=0.5$.
}
 \label{fig:phaseDigaram}
 \end{figure*}

Many efforts subsequently have been made to understand the above discrepancies.
Zucker~\cite{Zucker1986Production} systematically discussed three trust-producing modes: trust is tied to the past or expected exchange, to social characteristics, and to formal societal structures.  
Ref.~\cite{Buchan2002SwiftNA} explored the impact of culture and its relationship with indirect reciprocity~\cite{Nowak2005Evolution}.
There are other explanations where the emergence of trust and trustworthiness is attributed to further factors, such as social awareness~\cite{Mcnamara2008Evolution}, reputation~\cite{Xia2022costly}, information~\cite{Michael2013Information}, delayed~\cite{Michael2012Delayed} or partial information~\cite{Paul2018Expectations} etc.  
While network reciprocity is found to potentially promote many altruistic behaviors, such as cooperation~\cite{Nowak2006Five}, fairness~\cite{Page2000The}, and honesty~\cite{Lying2020Capraro}, Ref.~\cite{Kumar2020the} however reveals that its impact on the evolution of trust and trustworthiness is marginal for whatever types of underlying networks of the population. 
Note that, most of aforementioned studies follow the imitation learning rule~\cite{Roca2009Evolutionary}, such as the Moran or Fermi rule~\cite{Szabo1998Evolutionary}, where individuals imitate the strategies of their neighbors who may have higher payoffs. In essence, imitative learning can be taken as a simple version of social learning~\cite{Bandura1977social}, where individuals learn from others in their socio-economic activities, through observations or instructions, which may or may not involve direct experiences.

While social learning is ubiquitous in nature and our society, there is however a different paradigm of learning that has been largely ignored --- the reinforcement learning (RL)~\cite{Sutton2018reinforcement}. 
As one of the main classes in machine learning algorithms, RL specializes in decision-making based upon experience and has achieved
tremendous success in many aspects of science and technology, especially after the marriage to deep learning~\cite{Lecun2015deep}. 
Yet, only recently, RL starts to be applied to the evolutionary game theory to help understand the emergence of cooperation~\cite{Zhang2020understanding, Song2022reinforcement, Ding2023emergence}, the resource allocation~\cite{Andrecut2001q, Zhang2019reinforcement}, and other collective behaviors in complex systems~\cite{Zhang2020oscillatory, Tomov2021multi, He2022q, Shi2022analysis}.

In fact, the RL has a solid foundation in neuroscience and the existing evidences have shown that the working area and the physiological processes for RL are different from those of social learning, manifesting itself as a fundamentally distinct learning paradigm~\cite{Lee2012neural, Rangel2008framework, Olsson2020neural}. 
In the social learning paradigm, the observations of utility are first made, and then their decisions are made by utility comparison with others, e.g. imitating the strategies of those peers who have higher utilities. As a kind of value-based learning, individuals score different actions in RL and action is chosen probabilistically based on these scores.
Here, the most important distinction is that social learning is based upon the utility-comparison rule within its neighborhood, while each individual with RL develops a unique policy by self-reflection. The rule remains the same in social learning, but the policies in RL are coevolving with their surroundings and could be unique for each player. This makes the two learning paradigms fundamentally different.
Within the paradigm of RL, we are interested in the following questions: 
\emph{Is such endogenous self-reflective learning way capable of providing an new paradigm to decipher the emergence of trust and trustworthiness?} It could provide fundamentally different insights from the current social learning paradigm by exogenous comparison with peers.

In this work, we investigate the evolution of trust and trustworthiness within the paradigm of reinforcement learning. 
Specifically, we adopt a Q-learning algorithm~\cite{Watkins1989learning, Watkins1992Q} to study the trust game, where each person is guided by two Q-tables, respectively for the role of trustor and trustee.
In the two-person scenario, we find that a high level of trust and trustworthiness emerges when individuals both respect the historical experience and have a long-term vision. Similar phenomena are observed when the two-person scenario is extended to a population level. The analysis of the Q-tables reveals underlying mechanism behind where a crossover of individuals' preferences is detected. We also compute the phase diagram within the associated game parameters, and several boundaries are identified for the onset of the emergence and for regions of different prevalences. 

The remainder of this work is organized as follows:
we introduce our Q-learning setup for the trust game in Sec. 2\ref{sec:model}.
In Sec. 3\ref{sec:results}, we show results for the two-person scenario, and provide a mechanistic analysis.
The impact of gain factor on trust evolution is studied, and the boundary analysis is conducted in Sec. 4\ref{sec:appreciation}.
In Sec. 5\ref{sec:lattice}, we extend our study of the evolution of trust within a population on a 1d lattice.
Finally, we conclude our work together with discussions in Sec. 6\ref{sec:discussion}.
%-------------------------------------------------------------------------------------------------------------------------%
\section{2. Methods}\label{sec:model}
%\subsection{2.1. The trust game}

We start by introducing the trust game~\cite{Berg1995Trust}, where two agents are engaged and one acts as a trustor and the other as a trustee. Initially, the trustor is endowed with one monetary unit, who can choose either to stay the status quo or to invest the trustee with trust by transferring an investment fraction $x\in(0,1]$. With the status quo, denoted as not trust (N), the game is over. With trust (T), the investment is multiplied by a gain factor $g$, the game enters into the second stage. The trustee has to make a choice between reciprocity (R) and betrayal (B). In the former, the trustee transfers a return fraction of its earnings $\omega\in(0,1]$ to the trustor as return, i.e. $\omega gx$. The choice of betrayal means no return, the trustee walks away without reciprocity. At the end of the second session, the final payoffs for the trustor and trustee are, respectively, $1-x+\omega gx$ and $(1-\omega)gx$ if R is chosen, whereas the payoffs correspond to $1-x$ and $gx$ if B is selected. Here, the investment of the trustor is a manifestation of trust, and the transfer back to the trustor can be interpreted as trustworthiness. Table~\ref{tab:payoff} summarizes the payoff matrix for the two agents.

\begin{table}[]
\resizebox{72mm}{9mm}{
\begin{tabular}{c|ccc}
\toprule [1.0pt]
\hline
\diagbox{$i$ (trustor)}{$j$ (trustee)} & R & \quad B\\
\midrule [0.5pt]
T & ${1+(g\omega-1)x, (1-\omega)gx}$ & \quad ${1-x, gx}$ \\
N & ${1, 0}$ & \quad ${1, 0}$ \\
\bottomrule[1.0pt]
\end{tabular}}
\caption{The payoff matrix in the trust game. Within each item, the first payoff is for the trustor (row player) and the second is for the trustee (column player).}
\label{tab:payoff}
\end{table}

In the one-shot anonymous scenario, it's obvious that the trustee prefers to walk away, and betrayal is the reasonable choice to maximize its payoff. Likewise, the trustor is supposed not to trust as no return is expected. Therefore, the rational solution for the trust game ends with the solution (N, B).  Notice that, from the payoff structure in Table~\ref{tab:payoff}, the trust game bears a strong resemblance to the Prisoner's Dilemma, but the two are fundamentally different because the two agents in the latter are symmetrical, while they play different roles and act in sequence in the trust game. 

For simplicity, we stick to this 2-player scenario and study their temporal evolution, where the two players play the role of trustor and trustee in turn. Specifically, they adopt a Q-learning algorithm~\cite{Watkins1989learning, Watkins1992Q}. In this algorithm, each player has two Q-tables denoted as $Q^A_{s_n}$ and $Q^B_{s_n}$ in hands that guide their decision-making for the role of trustor and trustee, respectively, see Table~\ref{tab:table}. 
One takes an action from the action set $\{T,N\}$ as a trustor, or from $\{R, B\}$ as a trustee.
The system is in one of four states denoted as $\mathbb{S}=\{s_1, ..., s_{4}\}$ with $s_1=TR, s_2=TB, s_3=NR, s_4=NB$ as shown in $Q^A_{s_n}$, whereas the states in $Q^B_{s_n}$ are the same but just reshuffled, $s_1=RT, s_2=RN, s_3=BT, s_4=BN$. The element $Q_{s, a}$ in the Q-tables is a value function used to measure the value of action $a$ in the given state $s$, which is constantly updated over time. Players aim to find the optimal policy by Q-learning in terms of maximizing the expected accumulated reward.

Without loss of generality, two players are initially assigned with a random strategy from $\{$TR, TB, NR, NB$\}$ with equal probabilities. Following the common practice, the two roles are switched in turn for the two players from round to round, though the way of randomly assigned role in each round does not change our findings.  
In round $t$, with a probability $\epsilon$, two players independently choose an action $a_t$ at random from the corresponding action set to conduct a trial-and-error exploration. Otherwise, they choose the action $a_t$ with the larger Q value within the given state $s_t$ in the associated Q-table. This completes the stage of decision-making for their actions. Afterwards, each player tries to draw some lessons by revising the corresponding action-state value that has been adopted in this round, i.e. the element $Q_{s_t,a_t}$ in the Q-table. The update is as follows
\begin{equation}
\begin{aligned}
Q_{s, a}(t+1) &=Q_{s, a}(t)+\alpha\left(r+\gamma \max _{a^{\prime}} Q_{s^{\prime}, a^{\prime}}(t)-Q_{s, a}(t)\right) \\
&=(1-\alpha) Q_{s, a}(t)+\alpha\left(r+\gamma \max _{a^{\prime}} Q_{s^{\prime}, a^{\prime}}(t)\right),
\end{aligned}
\end{equation}
where $s, a$ represents the current state $s_t$ and action $a_t$ of the focal individual, and $s^{\prime}, a^{\prime}$ are the state $s_{t+1}$ and action $a_{t+1}$ at $t+1$. $r$ is the reward one obtained for the action $a_t$ within state $s_t$, with reference to Table~\ref{tab:payoff}. 
$\alpha\in(0,1]$ is the learning rate, which captures the contribution of current step. A larger value of $\alpha$ means that the agent is more forgetful as old Q values tend to be more rapidly modified. $\gamma\in[0,1]$ is the discount factor, measuring the weight of future rewards, as $\max _{a^{\prime}} Q_{s^{\prime}, a^{\prime}}(t)$ is the maximal value expected in the new state. This completes the stage of Q-table updating, and a single round is done. The evolution protocol is summarized in Fig.~\ref{fig:protocol} in Appendix~\ref{sec:appendixA}A for clarity.

\begin{table}[]
\resizebox{75mm}{14mm}{
\begin{tabular}{c|cc c|cc}
\arrayrulecolor{tabcolor}\toprule [1.4pt]
\hline
\diagbox{State}{Action} & T ($a_{1}$) & N ($a_{2}$)&  \quad \diagbox{State}{Action} & R ($a_{1}$) & B ($a_{2}$) \\
\midrule [0.5pt]
\hline
(T,R) ($s_{1}$) & $Q_{s_{1}a_{1}}$ & $Q_{s_{1}a_{2}}$ & \quad (R,T) ($s_{1}$) &$Q_{s_{1}a_{1}}$ & $Q_{s_{1}a_{2}}$ \\
\rowcolor{gray!40}(T,B) ($s_{2}$) & $Q_{s_{2}a_{1}}$ & $Q_{s_{2}a_{2}}$&\quad (R,N) ($s_{2}$) \quad & $Q_{s_{2}a_{1}}$ & $Q_{s_{2}a_{2}}$ \\
(N,R) ($s_{3}$) & $Q_{s_{3}a_{1}}$ & $Q_{s_{3}a_{2}}$ & \quad (B,T) ($s_{3}$) & $Q_{s_{3}a_{1}}$ & $Q_{s_{3}a_{2}}$ \\
\rowcolor{gray!40}(N,B) ($s_{4}$) & $Q_{s_{4}a_{1}}$ & $Q_{s_{4}a_{2}}$& \quad (B,N) ($s_{4}$) & $Q_{s_{4}a_{1}}$ & $Q_{s_{4}a_{2}}$\\  
\hline
\bottomrule[1.4pt]
\end{tabular}}
\caption{The two Q-tables for each individual in the 2-player scenario, $Q^A_{s_n}$ (left) and $Q^B_{s_n}$ (right) are Q-tables respectively for the role of trustor and trustee.}
\label{tab:table}
\end{table}

\begin{figure*}[htbp]
\centering
\includegraphics[width=0.4\linewidth]{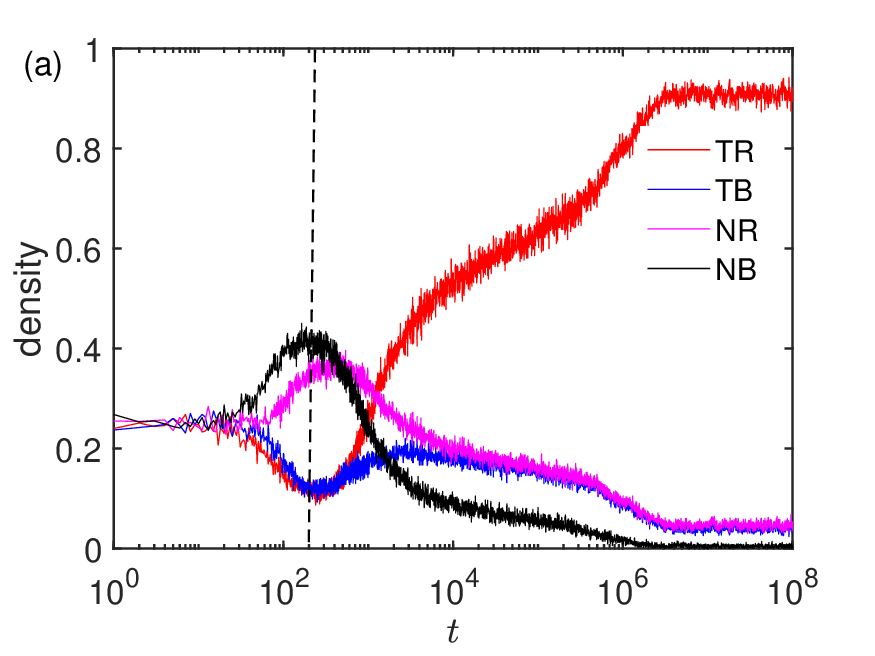}
\includegraphics[width=0.4\linewidth]{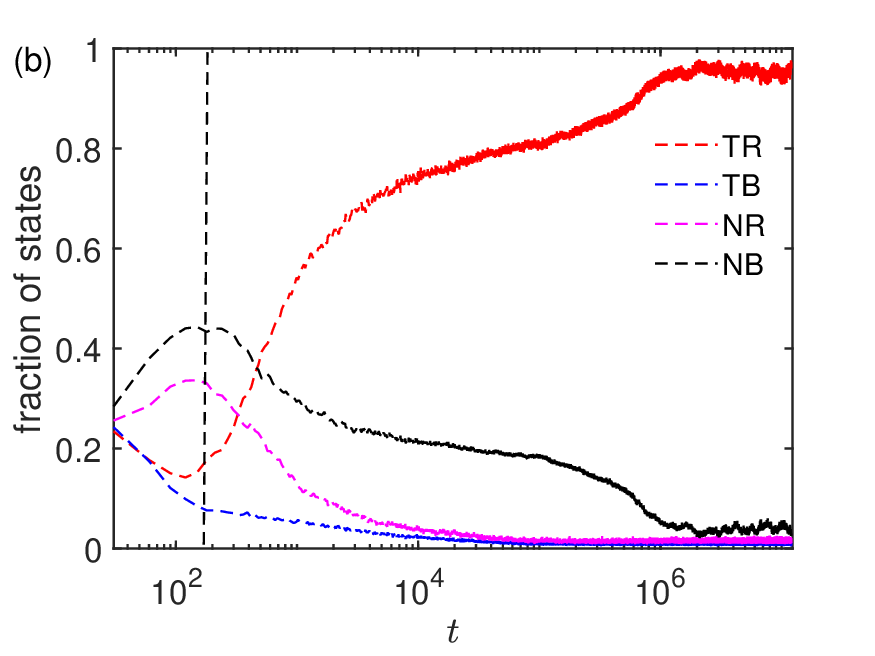}\\
\includegraphics[width=0.4\linewidth]{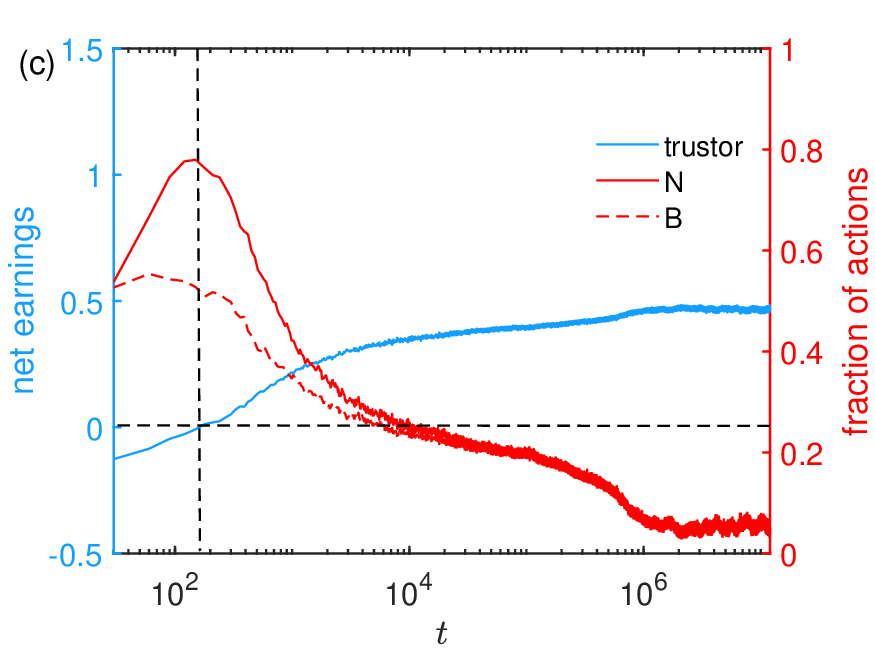}
\includegraphics[width=0.4\linewidth]{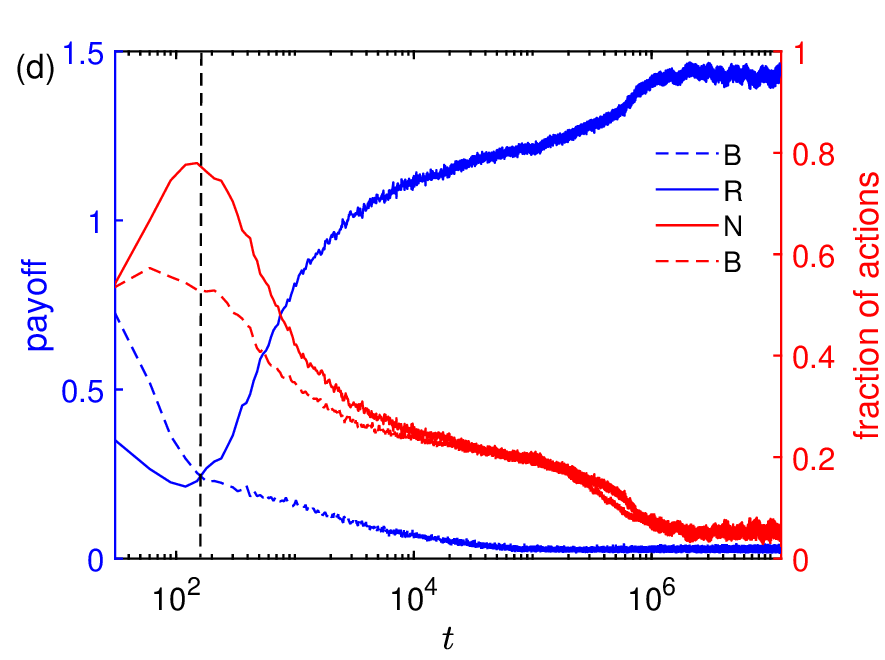}
\caption{
%(Color online) 
\textbf{Crossover in time series.}
Typical time series in the 2-player scenario for the learning parameter combination $\alpha=0.1$ and $\gamma=0.9$.
(a) The time series for the fractions of four strategy combinations;
(b) Four time series at the pairwise level, by simultaneously monitoring the fractions of the two players' strategies;
(c) The evolution of net payoffs to the trustor;
(d) The evolution of payoffs corresponding to different actions chosen by the trustee. 
The fractions of no trust (N) and betray  (B) are provided in (c, d) for reference, with the corresponding $y$-axis being put at the right side. 
The vertical dashed lines mark the same transition moment where the level of trust and trustworthiness starts to rise.
Each data is averaged over 500 realizations in (a-d), besides a sliding window average of 60 steps is conducted in (c, d).
Other parameters: $\epsilon=0.01$, $g=3$, $x=1.0$, and $w=0.5$.
}
\label{fig:ts}
\end{figure*}

In our practice, we focus on the case with fixed game parameters $g=3$, $x=1.0$, and $\omega=0.5$~\cite{Kumar2020the}. This corresponds to such a scenario: the trustor transfers all the money to trustee, and after the money is tripled, half of the profit is returned to the trustor.  
%Unless stated otherwise, the exploration probability $\epsilon$  is chosen to be 0.01, to balance the random exploration and the exploitation of the Q-table. 
Note that, we adopt a time-varying form for both learning rate and exploration rate at the beginning~\cite{Auer2002finite}, i.e. $\epsilon(t)=\max \{ \frac{4.0}{\sqrt{t}}, \epsilon\}$, $\alpha(t)=\max\{\frac{6.0}{\sqrt{t}}, \alpha\}$. This means that the two values starts with a large value, but as time goes by they get fixed once approach the desired value ($\epsilon$ and $\alpha$ will be set in the following studies), see Fig.~\ref{fig:deltaQ}(c). Though this setup does change the evolutionary outcome in the long term compare to the case for all-fixed parameters. Details see Appendix~\ref{sec:appendixA}B. 
For each case, the simulation is run over $10^8$ time steps to guarantee the evolution reaches a stable state.

%-------------------------------------------------------------------------------------------------------------------------%
\section{3. Results}\label{sec:results}
\subsection{3.1. Emergence of trust and trustworthiness}
We find that a prominent emergence of both trust and trustworthiness is observed in a physically meaningful region, see the phase diagram Fig.~\ref{fig:phaseDigaram}.
It reports that the final fractions of the four strategies $\{$TR, TB, NR, NB$\}$ in the learning parameters domain ($\gamma$, $\alpha$). As shown, there is a red region where the fraction of TR dominates (its fraction is larger than 0.6), where the learning rate $\alpha$ is small and the discount factor $\gamma$ is large. This observation means that when individuals focus on both historical experience and the long-term vision, both levels of trust and trustworthiness rise. Otherwise, either a forgetful property (a large $\alpha$) and/or a short-term vision (a small $\gamma$) lead to the failure of their emergence. Notice that, the fraction of the strategy TB remains small across the whole domain, meaning that once the trust is adopted, the agent also shows trustworthiness when acting in the role of trustee --- trustors never betray.  But once distrust is chosen, reciprocity and betrayal are equally likely to be chosen. 

\begin{figure*}[htbp]
\centering
\includegraphics[width=0.33\linewidth]{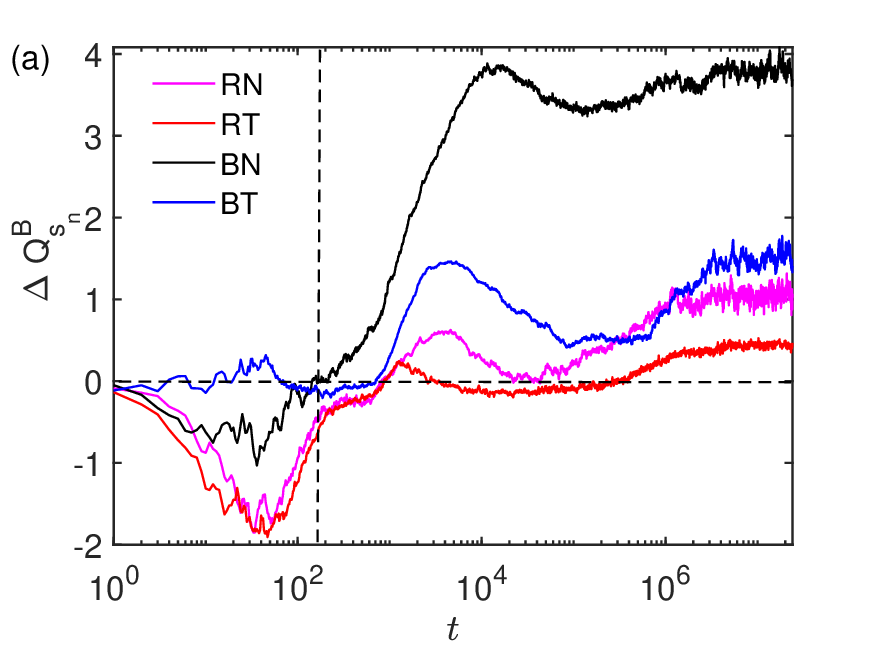}
\includegraphics[width=0.33\linewidth]{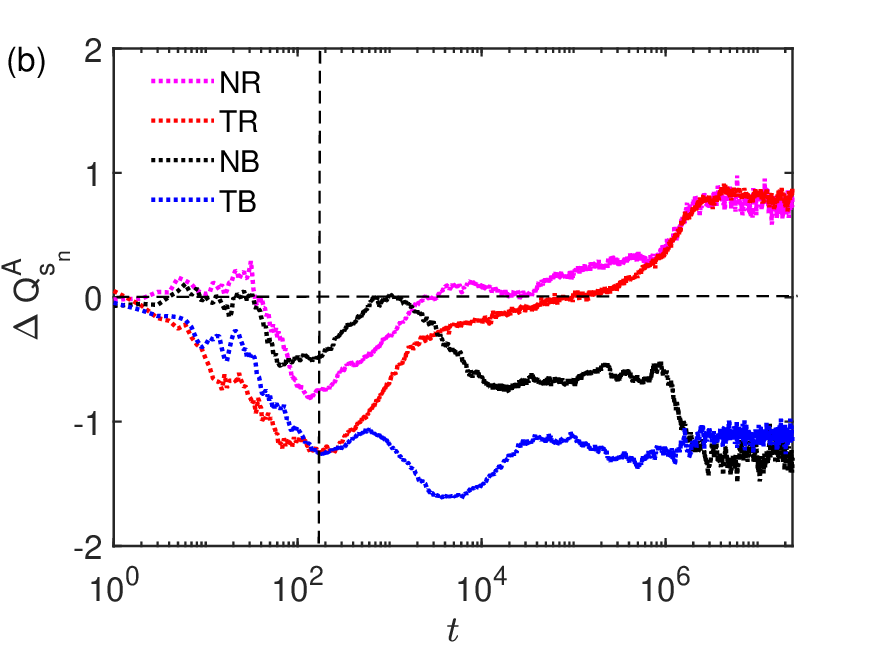}
\includegraphics[width=0.33\linewidth]{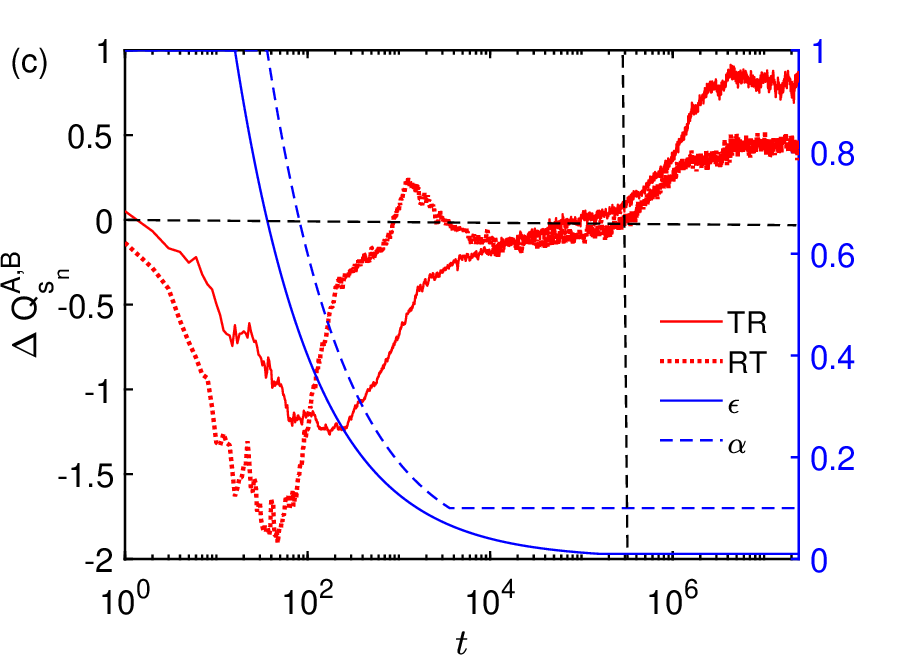}
\caption{
%(Color online) 
\textbf{Evolution of Q-tables.}
(a) The evolution of $\Delta Q^{B}_{s_n}$ for the trustee and (b) $\Delta Q^{A}_{s_n}$ for the trustor in all four possible states. 
If $\Delta Q^{A}_{s_n}>0$, the action of T is preferred within the state $s_n$ when in the role of trustor, otherwise N is flavored. Similarly, if $\Delta Q^{B}_{s_n}>0$ , the action R is flavored when acting as a trustee, otherwise B is preferred. The vertical dashed lines in (a) and (b) seperate the two stages, NB is preferred in the first stage, but is then followed by a turnover, where a positive feedback is formed to promote both the trust and trustworthiness prevalences.
(c) The two curves $\Delta Q^{A}_{TR}$ in (a) and $\Delta Q^{B}_{RT}$ in (b) are put together for clarity, along with the exploration rate $\epsilon$ and the learning rate $\alpha$. The vertical line in (c) marks the approximate transitions at which the individuals continue to choose to trust and show trustworthiness. Each data is averaged over 500 realizations.
Parameters: $\epsilon=0.01$, $g=3$, $\alpha=0.1$, $\gamma=0.9$. $x=1.0$, $w=0.5$.
}
 \label{fig:deltaQ}
 \end{figure*}
 
To understand the emergence, we show the time evolution of the four fractions for the case of $\alpha=0.1$ and $\gamma=0.9$, a typical combination of parameters located in the red region in Fig.~\ref{fig:phaseDigaram}, see Fig.~\ref{fig:ts}(a). We can see that their fractions start from around 0.25 due to random initialization, and then the two fractions of distrust (NR and NB) rise, but as time goes by these two fractions turn down, all three fractions except the strategy TR (the red line) decline, the fraction of NB (the black line) almost vanishes, and the system becomes stable in the end.

The reason for the initial decline in trust is straightforward since the level of reciprocity is low at the beginning (even below $50\%$), putting the trustor at a disadvantageous position, see Fig.~\ref{fig:ts}(b). The disadvantage of trustors is more clearly seen in Fig.~\ref{fig:ts}(c), where their net earnings are negative, being in a loss state. At the same time, the trustee is inclined to betrayal because as long as the trustor is willing to invest, betrayal tends to yield higher short-term payoffs compare to reciprocity. This is in line with the prediction from \emph{Homo economicus} perspective. This leads to a slight increase in the preference of betrayal, and consequently enhances the trend in the preference decline of trust. 

Unexpectedly, the fraction declines in trust ceases after around a hundred rounds and starts to turn up. 
This crossover can be understood by monitoring simultaneously the fractions of the two players' strategies [Fig.~\ref{fig:ts}(b)]. 
As can be seen, after the transient, the trustor tends not to invest in a trustee who has betrayed in the last round, but learns to invest to who showed the reciprocity, which explains the fraction of TB declines but TR starts to rise. 
This observation is also explained in Fig.~\ref{fig:ts}(d), where the advantage in the payoff by choosing B over R is lost at the moment as indicated by the dashed line. 
Intuitively, as the trustee learns that the action of betrayal incurs no investment, one starts to reciprocate the trustor instead of walking away. Once this trend starts, a higher level of reciprocity is preferred for the trustee, which in turn enhances the preference in action T for the trustor. This then forms a positive feedback that finally yields a high preference in both trust and trustworthiness.

%-----------------------------------------------------------%
\subsection{3.2. Evolution of Q-tables}

For a deeper understanding of the mechanism, let's direct our attention to the evolution of the two Q-tables. It shows the preference in NB like a \emph{Homo economicus} is only present at the very beginning, the evolution in the later stage aiming for maximizing payoffs in the long term forces them to turn to TR, which resembles human’s psychological changes. 

Specifically, we focus on the Q-value difference for each row of the two Q-tables, represented as $\Delta Q^{A,B}_{s_n}=Q^{A,B}_{s_{n}a_{1}}-Q^{A,B}_{s_{n}a_{2}}$, which determines the preferred action within the given state $s_n$ according to the idea of Q-learning. 
For example, if $\Delta Q^{A}_{s_n}\textgreater0$, this means that the action T is preferred within the state $s_n$ when acting as a trustor; otherwise N is supposed to be a better choice. 
Likewise, $\Delta Q^{B}_{s_n}\textgreater0$ means that action R is considered to be better when playing as a trustee, otherwise B is preferred. 
In our study, the evolution of the two Q tables for both individuals are found statistically the same, as their learning parameters are identical, implicating they have quite similar cognitive processes. 
Therefore, we only focus on the evolution of the $\Delta Q^{A,B}_{s_n}$ values for one of two individuals, as illustrated in Fig.~\ref{fig:deltaQ}.

Fig.~\ref{fig:deltaQ}(a) and~\ref{fig:deltaQ}(b) show respectively the time evolution of $\Delta Q^{B}_{s_n}$ and $\Delta Q^{A}_{s_n}$ for all four states. 
At the initial stage ($t\lesssim 100$, before the marked dashed line), $\Delta Q^{A,B}_{s_n}$ in both Q-tables mostly become more negative by learning, and the strategy for the individual converges to NB, where as a trustor one is unwilling to invest, and as a trustee one opts to betray. This is reasonable because betrayal for the trustee is better off than reciprocity in the short-term, since no investment can avoid potential money loss for the trustor. As a result, the dominating state in the system is NB for a trustor, and BN for a trustee, both players act indeed like a \emph{Homo economicus}.

As time comes to $t\approx 100$, the advantage of betrayal over reciprocity diminishes, because no investment comes from the trustor. This then causes a reversal of $\Delta Q^{B}_{BN}$ to be positive (the solid black line), and the action R is then preferred. This critical transition, however, does not immediately leads to the boom of trust or trustworthiness, because all four $\Delta Q^{A}_{s_n}$ in Fig.~\ref{fig:deltaQ}(b) at the moment are all negative, meaning that distrust is still dominating. Actually, the action of reciprocity is unstable, since $\Delta Q^{B}_{BN}>0$ and $\Delta Q^{B}_{RN}<0$, the state for the trustee oscillates between RN and BN. Therefore, at this  stage, still no driving force towards either trust or trustworthiness is seen, as shown in  Fig.~\ref{fig:deltaQ}(c), where TR is unstable for either trustor or trustee since the associated $\Delta Q^A_{TR}$ and $\Delta Q^B_{RT}$ are both negative at the left to the vertical dashed line.

An important change unfolds afterwards (i.e., $t\gtrsim 100$), as can be seen in Fig.~\ref{fig:deltaQ}(c). There, as the learning rate $\alpha$ and the exploration $\epsilon$ decreases, some optimal or nearly optimal policies have been learnt by individuals, and they rely more on their historical experience and conduct less random explorations. As shown, the trustee becomes gradually inclined to choose reciprocity rather than betrayal, as  $\Delta Q^B_{RT}$ stops decreasing and starts to grow. When this turnover is detected, the trustor also starts to trust as a response, where a turnover is also present for $\Delta Q^A_{TR}$, but with a time delay. 

Once both players turn to TR, this forms a positive feedback loop, both are well paid, leading to an increasing $\Delta Q$ that are both positive in the end.  
When the system becomes stable in the long run ($t\gtrsim 10^6$),  the trustee always chooses R since all its $\Delta Q^B>0$ [Fig.~\ref{fig:deltaQ}(a)]. On the trustor side, however, $\Delta Q^A_{s_n}>0$ only for the action of reciprocity (i.e. for $s_n=$ TR and NR), the trustor chooses not to trust  (for $s_n=$ TB and NB) when the trustee betrayed [Fig.~\ref{fig:deltaQ}(b)]. This punishment-like policy further forces the occasional betrayals back to the reciprocity. This then produces stable emergence of trust, where a decent level of trust and trustworthiness is seen.

Note that the final rise of $\Delta Q$ for both Q-tables also benefits from the diminishing exploration rate [Fig.~\ref{fig:deltaQ}(c)], where both trustor and trustee choose TR with a large probability. In a more noisy scenario (a large $\epsilon$), which can be interpreted as many misunderstandings or ``trembling hands"~\cite{Fundenberg1990evolution}, which can considerably suppress the level of the trust and trustworthiness, see Appendix C.

\begin{figure}[tbp]
 \centering
\includegraphics[width=1.0\linewidth]{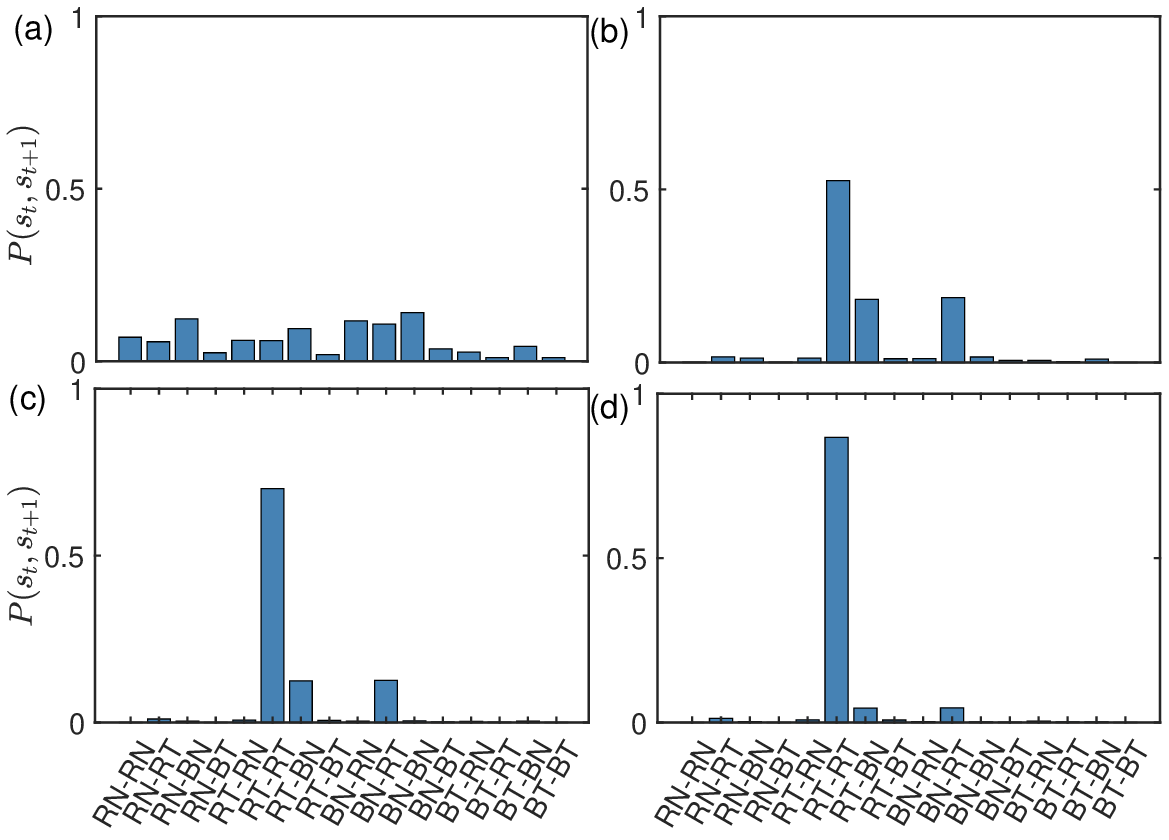}
\caption{
%(Color online) 
\textbf{Joint probabilities for two consecutive states.}
The joint probabilities $P({s_{t},s_{t+1}})$ of all sixteen state transitions in four typical stages: (a) $0 \sim 1000$, (b) $1.0\times10^3 \sim 1.0 \times10^5$, (c) $1.0\times10^5 \sim 1.0\times10^6$, (d) $2.0\times10^6 \sim 3.0\times10^6$ time steps. In the end, the system mainly stays in TR state. Each data is averaged over 500 realizations.
Parameters: $\epsilon=0.01$, $g=3$, $\alpha=0.1$, $\gamma=0.9$. $x=1.0$, and $w=0.5$.
}
 \label{fig:pdf}
 \end{figure}

The above analysis shows that the emergence of trust and trustworthiness is caused by the preference transition from NB to TR.
To further confirm the analysis, we compute the joint probability for two consecutive states $P({s_{t},s_{t+1}})$, all state transitions at different stages are shown in Fig.~\ref{fig:pdf}. In the early stage of evolution ($0\!<\!t\!<\!1000$), almost all mode transitions are detected, however, some modes are more likely to happen, such RN-BN, BN-RN, BN-BN, BN-RT, RT-BN, as shown in Fig.~\ref{fig:pdf}(a). This means that the state BN is the main state at this stage, in line with our above argument.
As analyzed above, once the trust and trustee start to form a positive feedback loop, the strategy of TR starts to dominate, as can be seen in Fig.~\ref{fig:pdf}(b). 
Apart from the dominating TR-TR mode, the other two bars BN-RT and RT-BN are also present, meaning that the strategy pair RT flips to BN from time to time, but the flipping back to RT is equally likely, as the two bars are of nearly the same height. 
In Fig.~\ref{fig:pdf}(c-d), the mode of RT-RT becomes even higher, though the other two bars (i.e. BN-RT and RT-BN) are still present due to the non-vanishing $\epsilon$. Till then, the evolution of trust and trustworthiness becomes stable.

Based on the above analysis, the emergence of trust can be roughly divided into three stages:
\begin{itemize}
\item [1)]
Initially, the trustee finds that betrayal is more profitable than reciprocating the trustor, and is thus inclined to choose B. Betrayal gradually becomes prevalent.
As a consequence, the trustor chooses to be non trusting as the net earning of investing is negative. The level of trust and trustworthiness both decrease.
\item [2)]
As less investment is detected, 	the tendency towards betrayal was reversed, the trustee starts to be inclined to reciprocating the investment. Once this turnover is on site, the trustor also starts to invest. 
\item [3)]
The preference change in TR for the two players then forms a positive feedback that strengthens the advantage of T and R in their Q-tables. In the end, the trustee prefers R in almost all scenarios, and the trustor trusts those reciprocating trustees but invests no money to the betrayed trustee as a punishment. This guarantees a decent level of trust and trustworthiness.
\end{itemize}

However, as the two learning parameters ($\gamma$, $\alpha$) deviate from the red region in Fig.~\ref{fig:phaseDigaram}(a), the mechanism behind stages 2) and 3) could be ruined. Analyses based on the evolution of Q-tables indicate that for a large $\alpha$, the past experience is rapidly washed out, so that no lesson can be drawn from the history. In this case, the evolution of the game degrades to the classic iterated scenario in the absence of Q-learning, where the trustor is not willing to invest, and trust and trustworthiness fail to emerge~\cite{Kumar2020the}. Meanwhile, for the case of small $\gamma$, the positive feedback between T and R fails to establish without confidence of the future reward. For a detailed analysis see Appendix~\ref{sec:failure}D. 

%-------------------------------------------------------------------------------------------------------------------------%
\section{4. Impact of gain factor and boundary analysis}\label{sec:appreciation}
\begin{figure}[tbp]
 \centering
\includegraphics[width=0.95\linewidth]{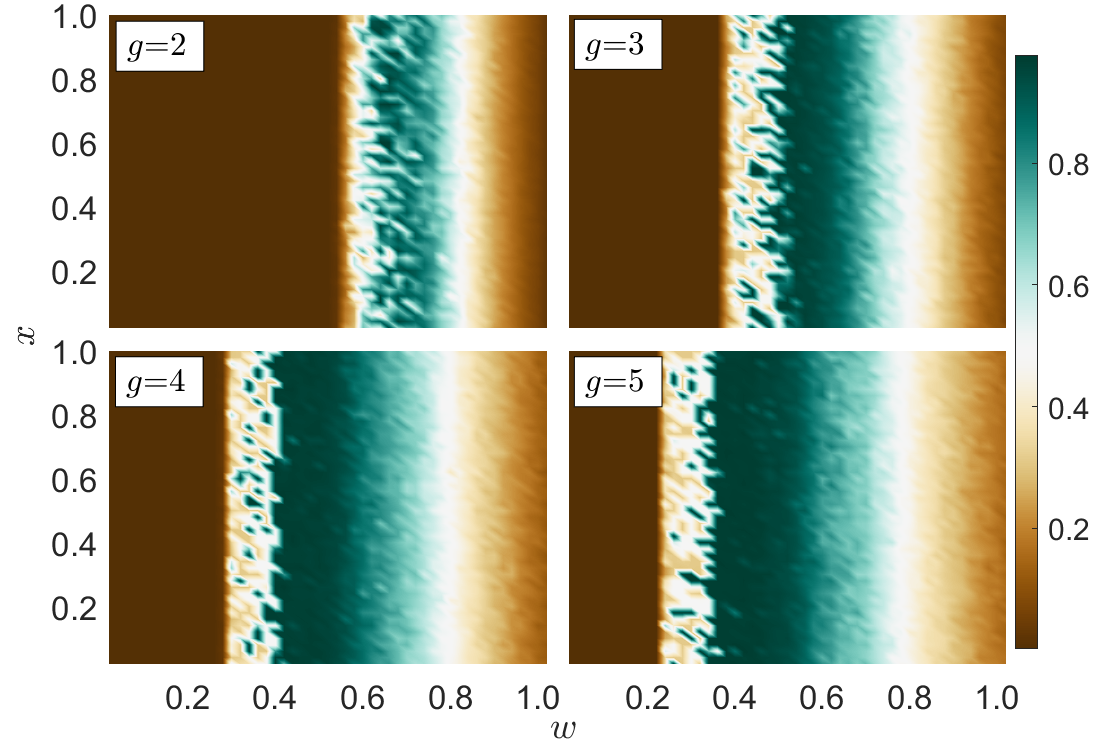}
\caption{
%(Color online) 
\textbf{Impact of gain factor.}
The color-coded fraction of the strategy TR in the parameter domain of $x-\omega$, where the gain factor $g=2,3,4$ and 5, respectively in (a-d). Each data is averaged $2\times10^4$ times after a transient of $2\times10^7$.
Other parameters: $\epsilon=0.01$, $\alpha=0.1$, and $\gamma=0.9$.
}
 \label{fig:appreciation}
 \end{figure}

The revealed mechanism is robust against the game parameter $x$, but shows intricate dependence on the return fraction $\omega$ and the gain factor $g$.
In the original model~\cite{Berg1995Trust}, the investment is tripled on the trustee side, while in some other work the gain factor $g$ is set to be 2 or other values~\cite{Gereke2018Ethnic,Woolley2017A}.
To systematically investigate the impact of the gain factor $g$ on the emergence of trust and trustworthiness, here we first show the dependence of the TR fraction on the investment fraction $x$ and the return fraction $\omega$ for $g=2, 3, 4, 5$, shown in Fig.~\ref{fig:appreciation}. As can be seen, the fractions of TR show no any dependence on the investment fraction $x$ in all four cases, which is reasonable since the investment fraction only affects the absolute payoff, but not the relative values in the Q-tables and thus the level of trust and trustworthiness shows no dependence on $x$. However, a larger gain factor widens the region where the trust and the trustworthiness emerge. We find that there is a simple relationship determining the left boundary. For a trustor, the bottom line to invest is not to lose money, conditioned by $gx\omega \geq x$, leading to the following inequality 
\begin{equation}
\begin{aligned}
g\omega \geq1.
\end{aligned}
\label{eq:boudnary}
\end{equation}
This immediately gives the left boundaries in Fig.~\ref{fig:appreciation}, i.e. $\omega_c^{(1)}=1/g$, which is well confirmed and explains the independence on $x$.

\begin{figure}[tbp]
 \centering
\includegraphics[width=0.9\linewidth]{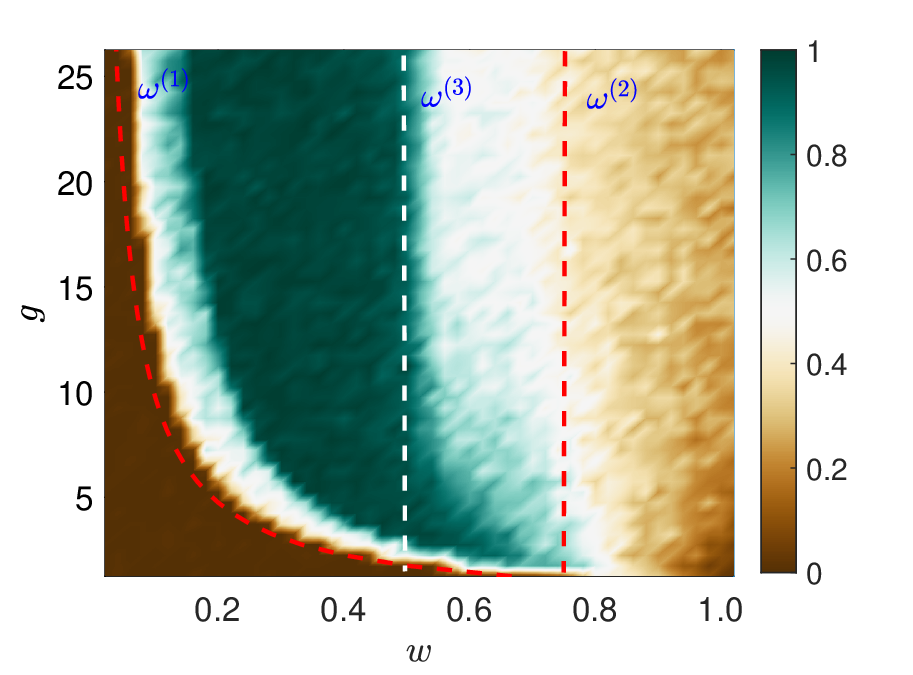}
\caption{
%(Color online) 
\textbf{Phase diagram and boundaries.}
The color-coded fraction of the strategy TR in the parameter domain of $\omega-g$, where $x=0.5$.
Three suggested thresholds (dashed line) are respectively corresponding to $\omega^{(1,3,2)}=1/g, 0.5$, and 0.8 from left to right.
Each data is averaged over 100 realizations.
Other parameters: $\epsilon=0.01$, $\alpha=0.1$, and $\gamma=0.9$.
}
 \label{fig:appreciationB}
 \end{figure}

This theoretical argument is better validated in Fig.~\ref{fig:appreciationB} within the parameter domain $\omega-g$ by fixing $x=0.5$. We see that the hyperbolic boundary fits perfectly the simulation results within a wide range of the gain factor $g$.
A closer lookup shows that the boundary slightly shifts to the right as $g$ increases compared with the theoretic prediction. This implies that when the gain becomes so high, the trustor would take some risk of losing money to invest the trustee.
Actually, the relationship revealed in Eq.~(\ref{eq:boudnary}) is supported by previous experiments~\cite{Anthony2011Elements,Lenton2011Incentivising,Johnson2011Trust}, where they found that as the gain factor increases, the promotion in trust is seen.

Interestingly, the right boundary seen in Fig.~\ref{fig:appreciation} seems independent on the gain factor $g$, where $\omega_c^{(2)}\approx 0.8$. This means that for a trustee, one would only show trustworthiness only if she can keep at least $20\%$ of the pie, otherwise one just walks away even if the gain factor is large enough that a positive earning is expected. This observation implies that the emergence of trustworthiness is built upon fairness,  individuals desire a fair division of the pie. 
Furthermore, $\omega^{(3)}=0.5$ seemingly marks as another threshold, below which a considerably high level of TR is then possible, especially for a large gain factor. This means that full reciprocity further requires that the trustee can keep at least half of the pie. 

 %--------------------------------------------------------------------------------------------%
\section{5. 1-dimensional Lattice} \label{sec:lattice}

\begin{figure}[tbp]
 \centering
\includegraphics[width=0.9\linewidth]{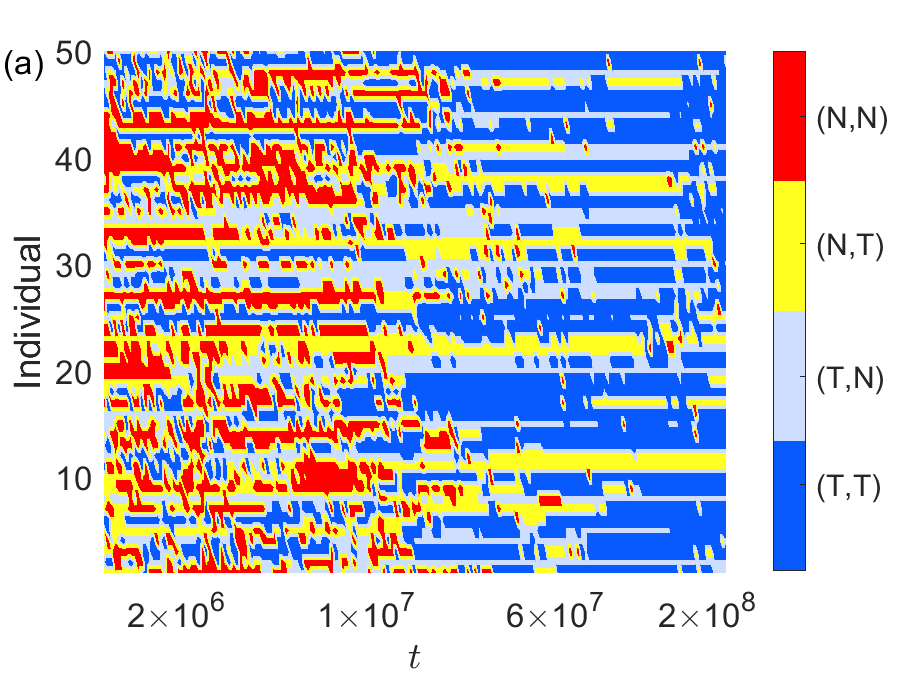}
\includegraphics[width=0.9\linewidth]{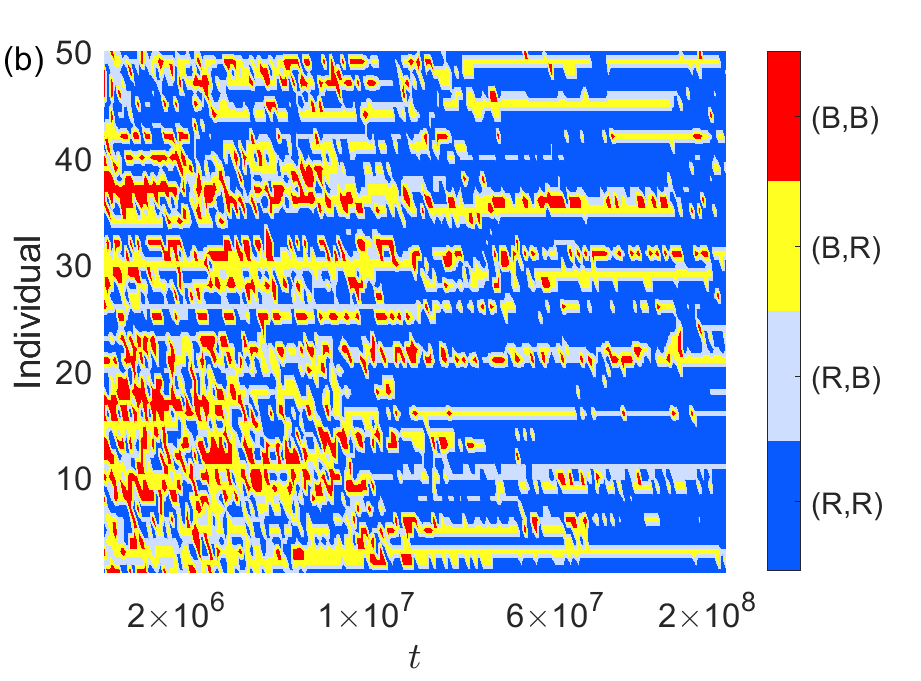}
\caption{
%(Color online) 
\textbf{Spatiotemporal evolution of 1d latticed population.}
 (a) and (b) correspond to scenarios where the individual $i$ acts as a trustor or a trustee, respectively. In each scenario, there are four action combinations, e.g., the strategy (N,T) in (a) means that as a trustor, the player chooses not to trust the left nearest neighbor, but trusts the right nearest neighbor.  
Parameters: $N=50$, $k=2$, $\epsilon=0.001$, $\alpha=0.1$, and $\gamma=0.98$.
}
 \label{fig:lattice}
 \end{figure}

In fact, the findings are not restricted to the above 2-player scenario, they are robust and can also be seen at the population level. An example of 1-dimensional lattice with the size $N=50$ and $k=2$ is shown in Fig.~\ref{fig:lattice}. Since there are two nearest neighbors for each individual in this scenario, the Q-table has to be expanded accordingly, detailed settings can be seen in Appendix~\ref{sec:lattice}E.
We find a qualitatively similar phenomenon that trust and trustworthiness tend to arise when both historical experiences and the long-term vision are emphasized, and the fraction of TR stabilizes at around 0.7 in the long run. 
In Fig.~\ref{fig:lattice}(a), we can see that at the early stage, the nontrusting strategy $(N,N)$ is dominating against their two nearest neighbors when acting as a trustor , but as time goes by, players gradually tend to invest. Also, the level of reciprocity is low at the beginning but rises to be high later on, see Fig.~\ref{fig:lattice}(b). These observations are similar to the 2-player scenario. Notice that, due to the presence of more neighbors, a smaller exploration rate $\epsilon$ and a higher expectation of future reward $\gamma$ are preferred to maintain a stable surrounding of a high level of TR, compared with the 2-player scenario.

Actually for a pair of players, they act according to the their two Q-tables regarding the other person, relying only on the agreement reached between them. 
Therefore, even when an individual has multiple neighbors, trust in one neighbor does not imply trust in others, i.e. the occurrence of trust and trustworthiness is by nature pairwise between the two involved individuals, independent of the rest. As a result, when extended to other complex networked populations, we expect that the mechanism and phenomena are similar to the 2-player scenario.%, which remains to be validated in the future.

 %-----------------------------------------------------------------------------------------%
\section{6. Discussion}\label{sec:discussion}
In summary, we have investigated the trust game within the paradigm of reinforcement learning, each player acts following a Q-learning algorithm, and we focus on the evolution of trust and trustworthiness. Surprisingly, high levels the trust and trustworthiness emerges in the two-player scenario when players both care about the historical experience and have long-term vision. The evolution of the associated two Q-tables reveals a crossover in the action preference, which resembles the psychological transition when we human beings playing the game. Our boundary analysis shows that a high level of trust and trustworthiness requires that the net earnings for the investment for the trustor is positive and the trustee can keep half of the earnings in hand. Furthermore, if the action choice deviates much from learnt Q-tables, this ``trembling hand" effect undermines the evolution of trust and trustworthiness, where the desired relationships are broken down. Finally, the emergence of trust and trustworthiness can also be seen when the scenario is changed into a latticed population. 

Interestingly, part of these observations were also seen in a series of experiments by Engle-Warnick and Slonim~\cite{Jim2004The,Engle2006Learning,Jim2001The}. They explored the evolution of strategies in repeated trust game experiments, and revealed that experience, attitude toward the future, institutions, and other factors have an important influence on the strategy selection in the repeated trust game. The results confirm that concerns for the future of repeated interactions and past experiences in game history are important for the persistence of trust.

Most importantly, we do not resort to any external factor as assumed in most previous work with social learning. Our reinforcement learning provides a natural explanation for the emergence of trust and trustworthiness. This indicates that past experience and the expectation for return in the future together as the endogenous factors are sufficient to trigger their emergence.   
In fact, existing efforts within this paradigm show that it can also provide explanations for understanding cooperation~\cite{Zhang2020understanding, Song2022reinforcement, Ding2023emergence}, resource coordination~\cite{Andrecut2001q, Zhang2019reinforcement} etc.  These work suggest that reinforcement learning demonstrates its power in explaining human behaviors, where the evolution of Q-table provides a uniform mechanism framework.
Given the consistent experimental deviation from the predictions of \emph{Homo economicus} regarding different altruistic behaviors~\cite{Camerer1995anomalies, Camerer2004advances, Henrich2005Economic, Johnson2011Trust}, such as cooperation, fairness, trust and trustworthiness, the reinforcement learning may provide a uniform paradigm to decipher complexities of human psychology, shedding new insights into the understanding of moral behaviors~\cite{Capraro2018grand, Capraro2021mathematical}.

Although we adopt reinforcement learning as our framework, we do not deny the value of social learning paradigm that has been widely used in previous studies. In fact, the two paradigms are not contradictory, but complementary to each other. 
Till now, there are plenty of experimental evidences in neuroscience showing that the decision making for both learning ways have solid neural bases~\cite{Lee2012neural, Rangel2008framework, Olsson2020neural},
and indicate that they may work for different scenarios. More probably, the learning processes in the real world are a mixture of the two when we human are dealing with complex issues.
An important question as the next step is to infer the type of learning from the behavioral experiments. Only when the learning paradigms in realities are clarified, we are on the right track to understand many important issues the human society such as cooperation, fairness, honesty, and so on. 

%-------------------------------------------------------------------------------------------------------------------------%
\section{Acknowledgments}
This work was supported by the National Natural Science Foundation of China [Grants Nos. 12075144,12165014]. ZGZ is supported by Excellent Graduate Training Program of Shaanxi Normal University [Grants No. LHRCTS23064].

\appendix
\section{Appendix A: Our protocol of Q-learning}\label{sec:appendixA}

\begin{figure}[tbp]
 \centering
\includegraphics[width=1.0\linewidth]{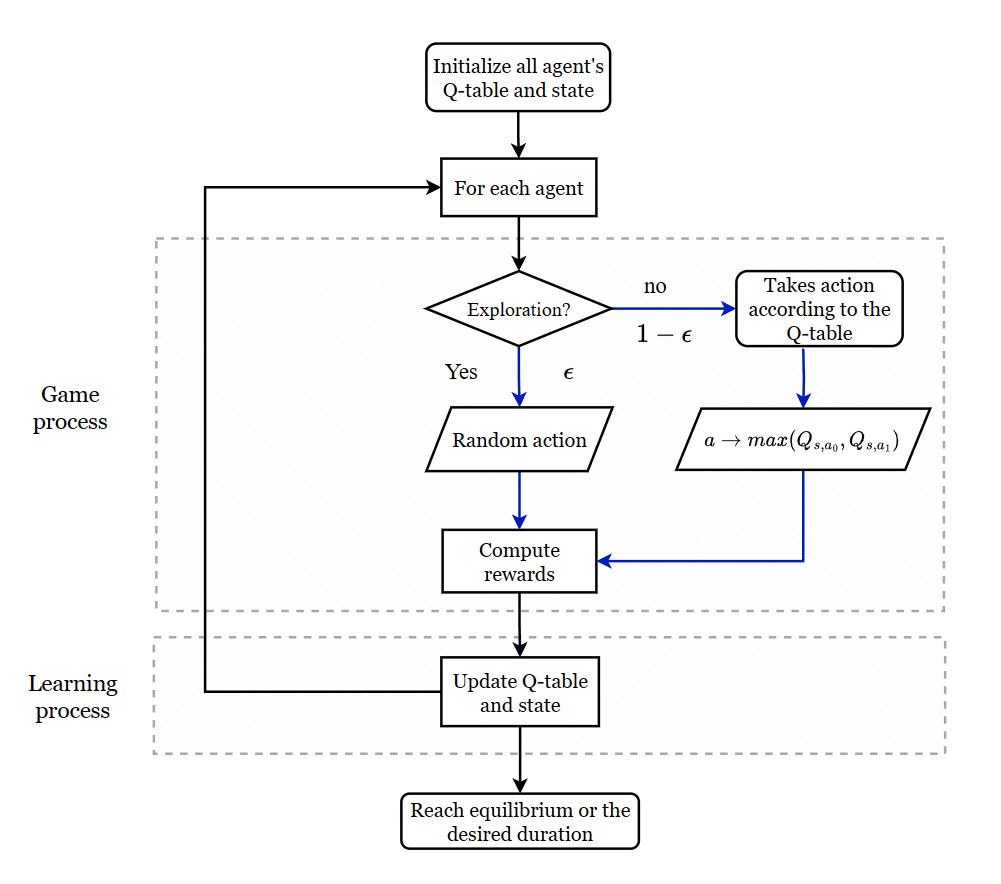}
\caption{
The protocol flowchart for the evolution of trust game.
}
 \label{fig:protocol}
 \end{figure}

For clarity, our Q-learning setup of trust game is summarized in Fig.~\ref{fig:protocol}, which is a synchronous updating protocol as follows:
\begin{itemize}
\item [1)]
 Initialize all Q-tables with small random number within $(0,1)$ to mimic the unawareness of agents to the environment,  and also a random strategy within $\{TR, TB, NR, NB\}$ for each agent. 
\item [2)] In the game process, each agent chooses an action by either exploitation or exploration, afterwards their rewards are obtained by collecting payoffs. 
\item [3)] The learning process is through the update their Q-tables, and their states also need to be updated. 
 \end{itemize}
 Repeat steps 2) and 3) until the system reaches statistically stable or the desired time duration.

\section{Appendix B: Setup of exploration rate and learning rate}\label{sec:appendixB}

To speed up the evolution, we adopt a time-varying form for the exploration rate $\epsilon$ and learning rate $\alpha$ at the initial stage, which is a common practice~\cite{Auer2002finite}. To be specific, $\epsilon=\frac{4.0}{\sqrt{t}}$, $\alpha=\frac{6.0}{\sqrt{t}}$, and remained unchanged after dropping to the desired values (e.g. 0.01 and 0.1 in Fig~\ref{fig:ts}, respectively). The logic behind $\alpha=\frac{6.0}{\sqrt{t}}$ can be interpreted as that naive individuals at the beginning put more weight on the present and future's reward since they know that they have almost no experience initially; but as time goes by,  as more experiences have been learnt, they put more weight on the history with a decreasing $\alpha$ till to the desired learning rate. The same logic applies to the exploration rate $\epsilon$, where more explorations are conducted at beginning when no policy is formed, but the explorations decrease to a minimal level when optimal policies are approached. This practice is generally able to accelerate the convergence of the evolution, but does not change the system evolution in the long run.

\begin{figure}[tbp]
 \centering
\includegraphics[width=1.0\linewidth]{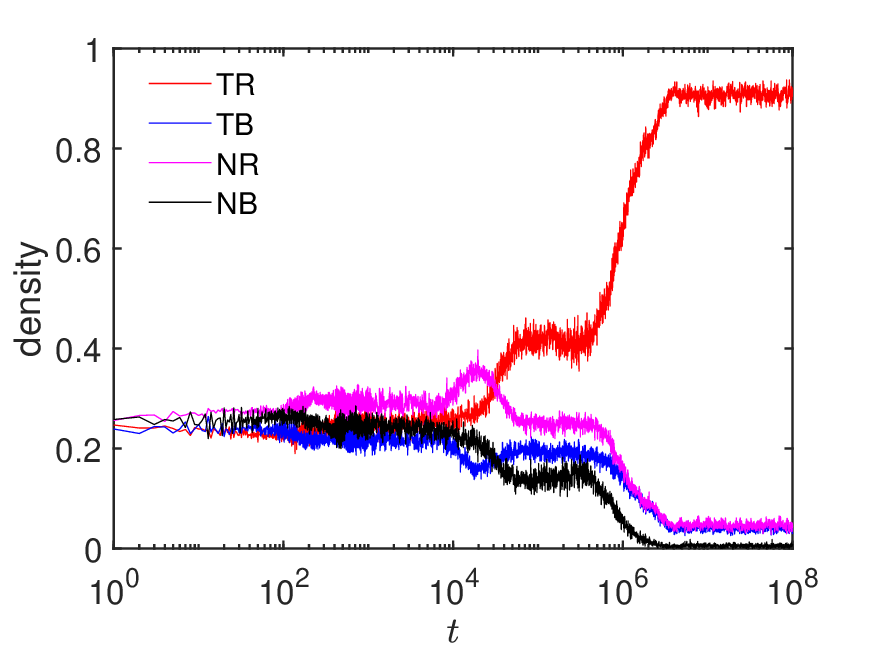}
\caption{
%(Color online) 
The evolution of four strategies fractions at an all-fixed exploration rate $\epsilon=0.01$ and all-fixed learning rate $\alpha=0.1$.
Other parameters and setup are exactly the same as Fig.~\ref{fig:ts}(a).
}
 \label{fig:all0.01}
 \end{figure}

As an comparison, Fig.~\ref{fig:all0.01} adopts the all-fixed the exploration rate $\epsilon$ and learning rate $\alpha$ all the time. It shows that the four fractions are consistent with the results in Fig.~\ref{fig:ts}(a) using the time-varying form, but indeed the transient takes a much longer time than the evolution in Fig.~\ref{fig:ts}(a).
%In addition, we have tried other values of $C$, its impact is marginal.

\section{Appendix C: The ``trembling hands" effect}\label{sec:appendixC}

\begin{figure}[tbp]
 \centering
\includegraphics[width=1.0\linewidth]{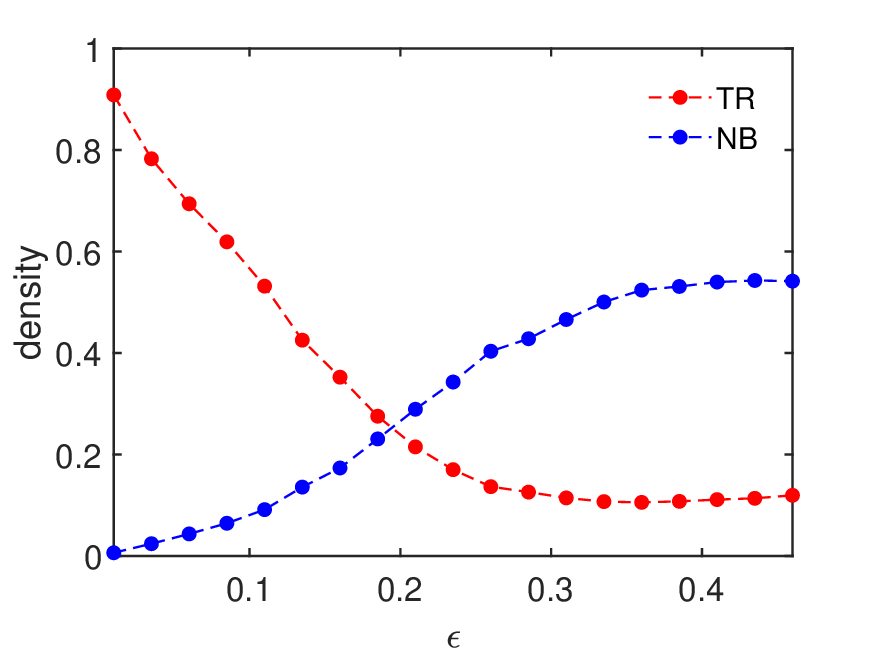}
\caption{
%(Color online) 
The ``trembling hands" effect in the 2-player scenario. The fractions of TR and NB versus the exploration rate $\epsilon$. Each data is averaged over 50 realizations.
Other parameters: $\alpha=0.1$, $\gamma=0.9$.
}
 \label{fig:trumblinghands}
 \end{figure}

The search of optimal policies benefits from the presence of the exploration rate $\epsilon$. But, once the evolution becomes stable, the players can be considered accumulated sufficient experiences that they have found their optimal policies. In that case, the exploration rate $\epsilon$ may be taken as the ``trembling hands"~\cite{Fundenberg1990evolution, Nowak2006Five}, players erroneously take actions, deviated from the guidance from their Q-tables. Fig.~\ref{fig:trumblinghands} shows that as the increase of $\epsilon$, the fraction of TR monotonically decreases, the fraction of NB rises accordingly. This suggests that a decent level trust and trustworthiness requires a weak “trembling hands” effect, too many misunderstandings would undermine and break down the foundation of trust.

\section{Appendix D: Failed cases of trust emergence}\label{sec:failure}
\begin{figure*}[tbp]
 \centering
\includegraphics[width=1.0\linewidth]{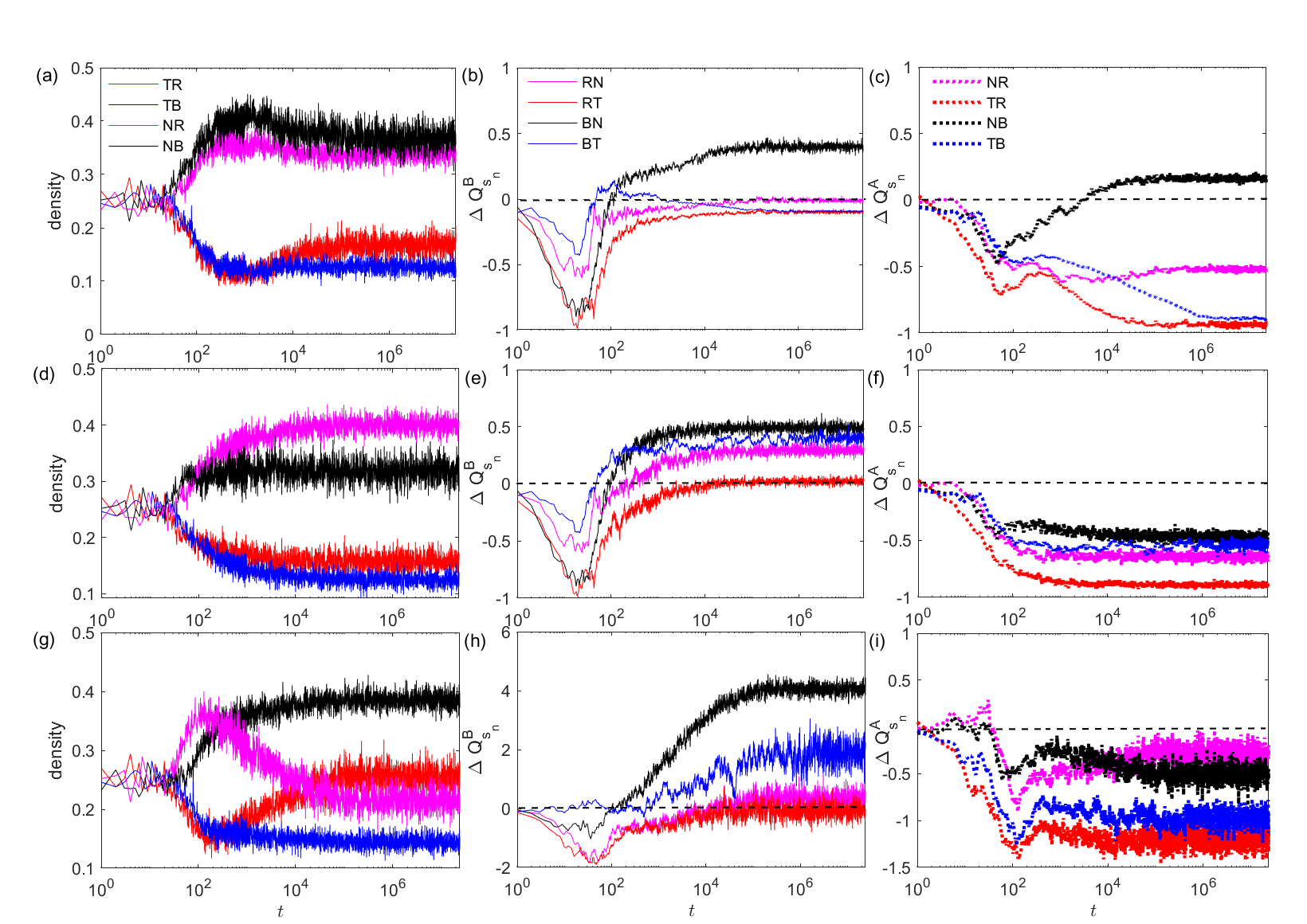}
\caption{
%(Color online) 
Three cases where the emergence of trust and trustworthiness fails. The parameter combinations are: (a-c) $\alpha=0.1$, $\gamma=0.1$; (d-f) $\alpha=0.9$, $\gamma=0.1$; (g-i) $\alpha=0.9$, $\gamma=0.9$. Other parameters are the same as Fig.~\ref{fig:deltaQ} 
}
 \label{fig:failure}
 \end{figure*}

As the two key learning parameters $(\gamma,\alpha)$ deviate the ideal combination, the emergence of trust and trustworthiness could fail. Three cases with typical parameter combinations are investigated, both the time series of the four fractions and their corresponding $\Delta Q_{s_n}^{A,B}$ are shown in Fig.~\ref{fig:failure}.
The failures can be attributed to two aspects. 

i) When the learning rate $\alpha$ becomes large [e.g. $\alpha=0.9$ in Fig.~\ref{fig:failure}(d-f) and (g-i)], this means that the historical experiences of players is removed immediately, almost no lesson is kept in the Q-table. In this scenario, the Q-learning algorithm loses its strength and the evolution degenerates to the traditional iterated trust game, where the trustor is not willing to invest in the trustee. Note that, once no investment is made, the reciprocating behaviors from the trustee is pointless, therefore, no decent level of trust and trustworthiness is seen, as shown in Fig.~\ref{fig:failure}(d) and (g). 

ii) When the discount factor $\gamma$ becomes smaller, this causes another problem. Without confidence of the future reward, it's hard for the trustor to foresee potential reciprocity from the trustee to select to trust, and vice versa. As a result, the positive feedback between trust and reciprocity fails to form. A decent level of trust and trustworthiness is still hard to see  [e.g. $\gamma=0.1$ in Fig.~\ref{fig:failure}(a-c) and (d-f)]. 

With these observations, it's reasonable to understand why a decent level of trust and trustworthiness is only observed for the parameter combination of small $\alpha$ and large $\gamma$ in Fig.~\ref{fig:phaseDigaram}.

\section{Appendix E: Setup of Q-table in 1D latticed population}\label{sec:lattice}

When we extend the 2-player scenario to the 1-dimensional latticed population, where each player connects 2 nearest-neighbors, i.e. the degree $k = 2$. The action set for the trustor is then extended to be $\mathbb{A}_r= \{(T, T), (T, N), (N, T), (N, N)\}$, where the first and the second are respectively the strategy playing against to its left and right nearest-neighbor.  Similarly, the action set for the trustee is $\mathbb{A}_e= \{(R, R), (R, B), (B, R), (B, B)\}$. The state for the trustor is expanded as $\mathbb{S}=\{s_1,...,s_{16}\}$, with $s_1=(TT, RR)$, $s_2=(TT, RB)$,..., $s_{16}=(NN, BB)$. For example, $s_2=(TT, RB)$ means that the player has invested her two nearest-neighbors in the last round, and the left one showed reciprocity but the right one betrayed. The two Q-tables $Q^{A,B}_{s_n}$ are illustrated in Table~\ref{tab:Ringtrustor} and Table~\ref{tab:Ringtrustee}, for the player who plays the role of trustor and trustee, respectively.

\begin{table}[]
\resizebox{73mm}{24mm}{
\begin{tabular}{c|c|c|c|c}
\toprule [1.2pt]
\hline
\diagbox{State}{Action} & (T, T) ($a_1$) & (T, N) ($a_2$) & (N, T) ($a_3$) & (N, N) ($a_4$)\\
\midrule [0.7pt]
(TT, RR) ($s_1$) & $Q_{s_1a_1}$ & $Q_{s_1a_2}$ & $Q_{s_1a_3}$ & $Q_{s_1a_4}$\\

\rowcolor{gray!40} (TT, RB) ($s_2$) & $Q_{s_2a_1}$ & $Q_{s_2a_2}$ & $Q_{s_2a_3}$ & $Q_{s_2a_4}$\\

(TT, BR) ($s_3$) & $Q_{s_3a_1}$ & $Q_{s_3a_2}$ & $Q_{s_3a_3}$ & $Q_{s_3a_4}$\\

\rowcolor{gray!40} (TT, BB) ($s_4$) & $Q_{s_4a_1}$ & $Q_{s_4a_2}$ & $Q_{s_4a_3}$ & $Q_{s_4a_4}$\\

(TN, RR) ($s_5$) & $Q_{s_5a_1}$ & $Q_{s_5a_2}$ & $Q_{s_5a_3}$ & $Q_{s_5a_4}$\\

\rowcolor{gray!40} ...  & ... & ... & ... & ...\\

(NT, RR) ($s_9$) & $Q_{s_9a_1}$ & $Q_{s_9a_2}$ & $Q_{s_9a_3}$ & $Q_{s_9a_4}$\\

\rowcolor{gray!40} ...  & ... & ... & ... & ...\\

(NN, RR) ($s_{13}$) & $Q_{s_{13}a_1}$ & $Q_{s_{13}a_2}$ & $Q_{s_{13}a_3}$ & $Q_{s_{13}a_4}$\\

\rowcolor{gray!40} ...  & ... & ... & ... & ...\\
(NN, BB) ($s_{16}$) & $Q_{s_{16}a_1}$ & $Q_{s_{16}a_2}$ & $Q_{s_{16}a_3}$ & $Q_{s_{16}a_4}$\\

\bottomrule[1.2pt]
\end{tabular}}
\caption{The Q-table for the trustor in the one-dimensional latticed population, where each player connects 2 nearest-neighbors.}
\label{tab:Ringtrustor}
\end{table}

\begin{table}[]
\resizebox{73mm}{24mm}{
\begin{tabular}{c|c|c|c|c}
\toprule [1.2pt]
\hline
\diagbox{State}{Action} & (R, R) ($a_1$) & (R, B) ($a_2$) & (B, R) ($a_3$) & (B, B) ($a_4$)\\
\midrule [0.7pt]
(RR, TT) ($s_1$) & $Q_{s_1a_1}$ & $Q_{s_1a_2}$ & $Q_{s_1a_3}$ & $Q_{s_1a_4}$\\

\rowcolor{gray!40} (TT, RB) ($s_2$) & $Q_{s_2a_1}$ & $Q_{s_2a_2}$ & $Q_{s_2a_3}$ & $Q_{s_2a_4}$\\

(RR, TN) ($s_3$) & $Q_{s_3a_1}$ & $Q_{s_3a_2}$ & $Q_{s_3a_3}$ & $Q_{s_3a_4}$\\

\rowcolor{gray!40} (TT, BB) ($s_4$) & $Q_{s_4a_1}$ & $Q_{s_4a_2}$ & $Q_{s_4a_3}$ & $Q_{s_4a_4}$\\

(RR, NT) ($s_5$) & $Q_{s_5a_1}$ & $Q_{s_5a_2}$ & $Q_{s_5a_3}$ & $Q_{s_5a_4}$\\

\rowcolor{gray!40} ...  & ... & ... & ... & ...\\

(RB, TT) ($s_9$) & $Q_{s_9a_1}$ & $Q_{s_9a_2}$ & $Q_{s_9a_3}$ & $Q_{s_9a_4}$\\

\rowcolor{gray!40} ...  & ... & ... & ... & ...\\

(BR, TT) ($s_{13}$) & $Q_{s_{13}a_1}$ & $Q_{s_{13}a_2}$ & $Q_{s_{13}a_3}$ & $Q_{s_{13}a_4}$\\

\rowcolor{gray!40} ...  & ... & ... & ... & ...\\
(BB, NN) ($s_{16}$) & $Q_{s_{16}a_1}$ & $Q_{s_{16}a_2}$ & $Q_{s_{16}a_3}$ & $Q_{s_{16}a_4}$\\

\bottomrule[1.2pt]
\end{tabular}}
\caption{The Q-table for the trustee in the one-dimensional latticed population, where each player connects 2 nearest-neighbors.}
\label{tab:Ringtrustee}
\end{table}

\bibliography{ref}
\end{document}